\documentclass[lettersize,journal]{IEEEtran}
\usepackage{amsmath,amsfonts}
\usepackage{algorithmic}
\usepackage{algorithm}
\usepackage{array}
\usepackage[caption=false,font=normalsize,labelfont=sf,textfont=sf]{subfig}
\usepackage{textcomp}
\usepackage{stfloats}
\usepackage{url}
\usepackage{verbatim}
\usepackage{graphicx}
\usepackage{cite}
\begin{document}
		%%%== TITLE AND AUTHOR DEFINITIONS GO HERE ==%%%
		%%\title{\large{\textbf{\uppercase{Entering the Metaverse: Enhancing User Access through 6G Wireless Systems}}}}
		
		\title{Mulsemedia Communication Research Challenges\\ for Metaverse in 6G Wireless Systems}
		%Please spell out first names and surnames
	\author{Ian F. Akyildiz, Hongzhi Guo, Rui Dai, and Wolfgang Gerstacker
	% <-this % stops a space
	\thanks{Ian F. Akyildiz is with Truva Inc., Alpharetta, GA 30022, USA. Hongzhi Guo is with the University of Nebraska-Lincoln, NE, 68588, USA. Rui Dai is with the University of Cincinnati, Cincinnati, OH, 45221, USA. Wolfgang Gerstacker is with the Friedrich-Alexander-Universität Erlangen-Nürnberg, Cauerstr. 7, D-91058 Erlangen, Germany}% <-this % stops a space
	%\thanks{Manuscript received April 19, 2021; revised August 16, 2021.}
}
		
		%%%=========%%%
\maketitle

\begin{abstract}
	Although humans have five basic senses, sight, hearing, touch, smell, and taste, most multimedia systems today only capture two of them, namely, sight and hearing. With the development of the metaverse and related technologies, there is a growing need for a more immersive media format that leverages all human senses. Multisensory media (Mulsemedia) that can stimulate multiple senses will play a critical role in the near future. This paper provides an overview of the history, background, use cases, existing research, devices, and standards of mulsemedia. Emerging mulsemedia technologies such as extended reality (XR) and Holographic-Type Communication (HTC) are introduced. Additionally, the challenges in mulsemedia research from the perspective of wireless communication and networking are discussed. The potential of 6G wireless systems to address these challenges is highlighted, and several research directions that can advance mulsemedia communications are identified. 
\end{abstract}


\section{Introduction}
The past several decades have seen significant progress in the development of multimedia communication and networking. Although the quality of video and audio has greatly improved, this has come at the cost of streaming high volumes of data through both wireless and wired networks. In recent years, the emergence of extended reality (XR), digital twins, digital currency, and Holographic-Type Communication (HTC) \cite{akyildiz2022holographic} has led to the development of the metaverse. However, the multimedia used in the metaverse only stimulates sight and hearing, neglecting three other fundamental human senses: touch, smell, and taste. For the metaverse to become a reality where users can live and work as if they were in the real world, all five human senses must be addressed and stimulated to create a fully immersive experience. This calls for a new format of multimedia that can incorporate multiple human senses.

\begin{figure*}[t]
	\centering
	\includegraphics[width=0.7\textwidth]{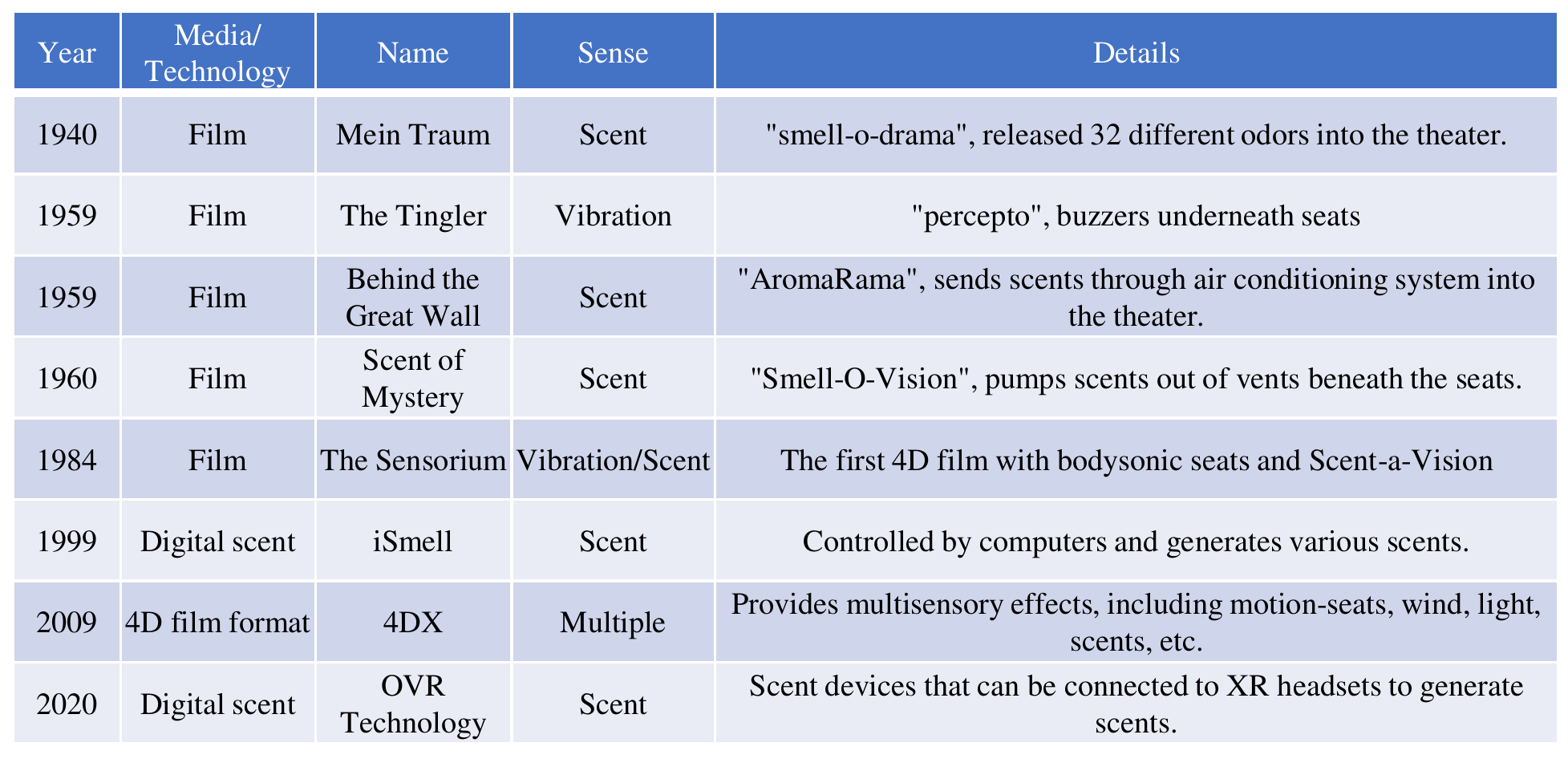}
	\vspace{-5pt}
	\caption{A brief summary of mulsemedia development milestones.}
	\label{fig:history}
	\vspace{-0pt}
\end{figure*}

Mulsemedia research emerged in 2010 \cite{ghinea2010user,kannan2010role,kannan2010digital} and it stands for media that includes three or more human senses \cite{ghinea2014mulsemedia}. Although this concept is relatively new, the employment of humans' senses beyond sight and hearing in film theaters started about one century ago. The invention of television in 1927 attracted many film customers. Film theaters started to come up with novel technologies to bring more customers back. The first film that adopted extra human senses is {{Mein Traum}} in the 1940s, which uses a so called {{``smell-o-drama''}} to release odors into the theater. However, this technology was not widely adopted due to a lack of mature support. Around 1960, more attempts appeared in film theaters, including the famous ``AromaRama'' and ``Smell-O-Vision'' which release scents through different systems into theaters. Also, vibration was employed beneath seats. After this, the first 4D film was screened in 1984, incorporating multiple humans' senses, such as vibration and scent. Nowadays, the 4DX format provides richer human senses for 4D films. Moreover, significant efforts were also spent to create high-quality media for other human senses besides sight and hearing. {{For example, digital scent technology has been available for a long time and some devices can be integrated with modern XR headsets.}} A brief summary of the mulsemedia development milestones is given in Fig.~\ref{fig:history}.       

In the past decade, researchers have developed various mulsemedia prototypes \cite{waltl2013end,marfil2022integration,covaci2022multisensory,bi2018dash} which use a range of senses, including scent, light, moisture, wind, haptic, and kinesthetic effects. To create mulsemedia content, annotators manually add information about wind speed, light intensity, and scent to videos, which are then played together at the client side. These pioneering works have proven the feasibility of mulsemedia communications, and results have shown that mulsemedia can significantly improve the Quality-of-Experience (QoE) \cite{yuan2014user}.

The metaverse \cite{wang2022survey} and the 6G wireless systems \cite{akyildiz20206g} have significantly changed the development of mulsemedia. First, the metaverse aims to provide seamless connections between the physical world and the virtual world, which demands truly immersive user experiences that can utilize more human senses beyond sight and hearing. This is the major driving application of mulsemedia communications. Second, the 6G wireless systems will provide ubiquitous connectivity with high data rates and low latency which enable mobile mulsemedia communications. Existing mulsemedia technologies are mainly used in 4D theaters due to complex sensing and display systems and high requirements on communication system performance. As a result, users have limited access to mulsemedia. The 6G wireless systems can dramatically change this situation by providing high-performance wireless communication for mulsemedia sensors, actuators and displays which can be embedded in computers and smart devices. 

It is important to understand the motivation of mulsemedia research, review the essential system blocks of mulsemedia communication, and identify the research challenges in the era of metaverse and 6G wireless systems. In this paper, we provide a comprehensive review of the past and the current status of mulsemedia communication, upon which we discuss its potential use cases and research challenges. In Section 2, we introduce the background of the metaverse which demands immersive mulsemedia communication, and we give an overview of the role of 6G wireless systems in the metaverse and mulsemedia communications. We also point out the relations between emerging technologies, such as HTC and XR which includes Augmented Reality (AR), Mixed Reality (MR), and Virtual Reality (VR). After that, we present the use cases of mulsemedia communications in Section 3. {{The mulsemedia communication system architecture and the technologies utilized at the sources, networks, and destinations are introduced in Section 4. In Section 5, we point out potential research opportunities in 6G wireless systems that can facilitate the development of mulsemedia communications.} Finally, this paper is concluded in Section 6.  
	
	\section{Background: Metaverse and 6G Wireless Systems}
	
	In this section, we introduce the background of mulsemedia, including its important role in the metaverse and the enablers of XR and HTC supported by the 6G wireless systems. 
	
	\subsection{Metaverse}
	
	{{The metaverse is defined as a hypothetical synthetic environment which is linked to the physical world \cite{lee2021all}. Users can interact with each other in the virtual environment provided by the metaverse \cite{dwivedi2022metaverse}.}} The metaverse is anticipated to become the next iteration of the Internet, consisting of immersive and hyperreal 3D digital virtual content. Technologies such as digital twins, avatars, and holograms enable the recreation and enrichment of the physical world in the metaverse. In addition to replicating physical environments, the metaverse allows for the creation of virtual information, objects, avatars, and environments that do not exist in the physical world. The integration of real and virtual content in the metaverse can greatly enhance user capabilities for work, entertainment, education, healthcare, and manufacturing \cite{tang2022roadmap}.  
	
	\begin{figure}[t]
		\centering
		\includegraphics[width=0.4\textwidth]{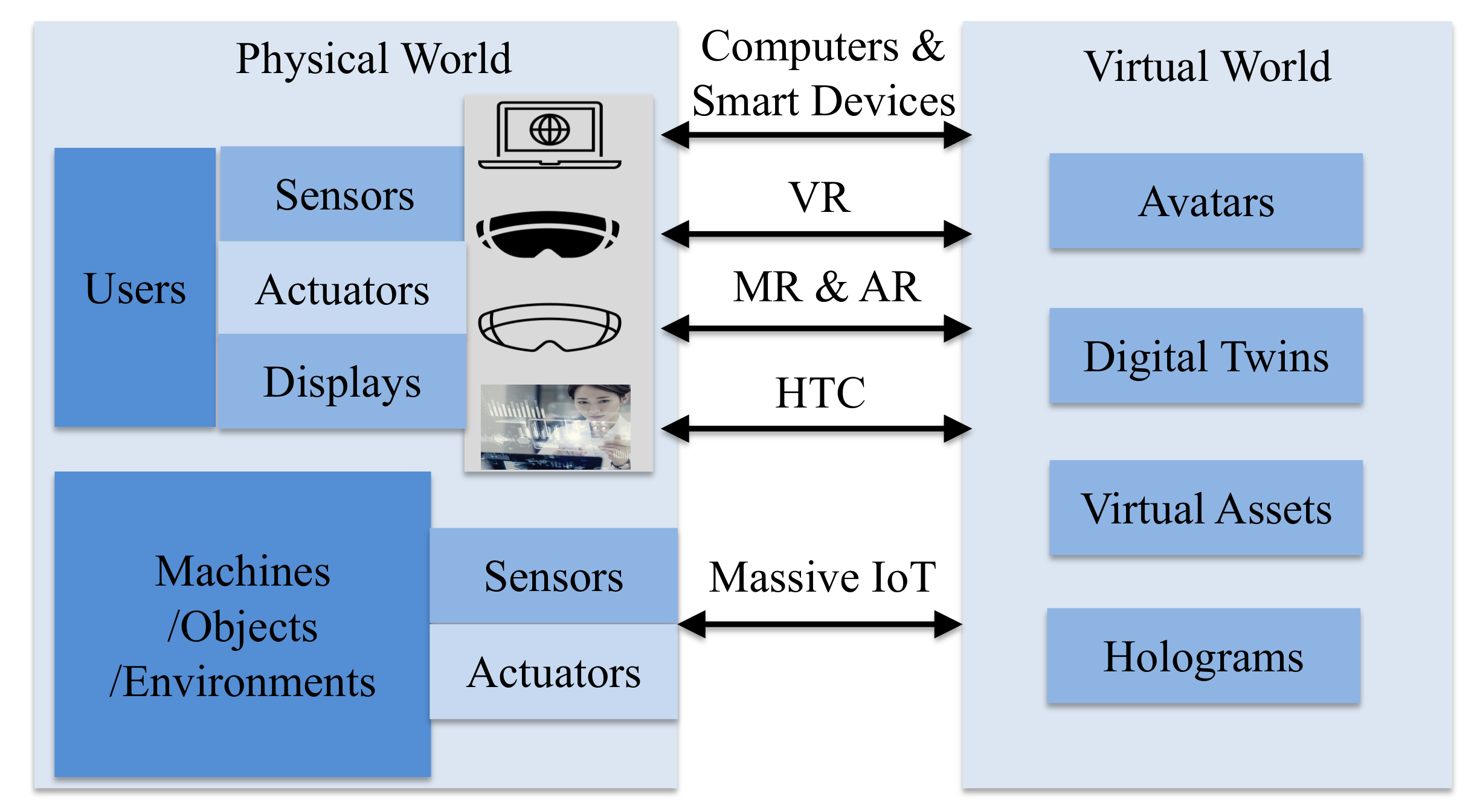}
		\caption{ The gateways that connect the physical world and the virtual world.}
		\label{fig:gateway}
	\end{figure}
	
	The metaverse will come in various formats, ranging from simple computer games to complex digital twins, enabling a plethora of novel applications, such as entertainment, gaming, industrial design and simulation, and more. {{Cryptocurrency also plays an important role in the metaverse. Virtual currencies and assets based on Non-Fungible Tokens (NFTs), blockchain, and Web 3.0 enable social economics in virtual worlds with high-security levels.} Moreover, blockchain and Web 3.0 provide fully distributed services and protect users' digital identities. 
		
		{{Since the metaverse focuses on social interactions, ethical issues need to be considered in its development. Users must be accountable for their activities in the metaverse \cite{fernandez2022life}.  }}
		
		In order to access metaverse content, users require various devices such as smartphones, tablets, computers, Head-Mounted Displays (HMDs), light field displays, and others, as shown in Fig.~\ref{fig:gateway}. Unlike existing multimedia, high-quality metaverse content is created in formats such as 360-degree videos, holograms, and mulsemedia, among others. These new media formats represent complex information and require high data rates, often as large as several gigabits per second \cite{akyildiz2022holographic}. Existing multimedia streaming protocols may experience issues such as network congestion, storage overload, and long latency, which can drastically reduce the Quality of Experience (QoE) for users. Additionally, content creation for the metaverse relies on the Internet of Everything (IoE), which remains a challenge in 5G wireless systems. While the Internet of Things (IoT) only networks physical smart devices, the IoE networks people, things, data, and processes to provide intelligence and cognition across networked environments. As a result, the IoE demands more comprehensive and ubiquitous network connectivity and intelligence than the IoT.

		\begin{figure}[t]
			\centering
			\includegraphics[width=0.47\textwidth]{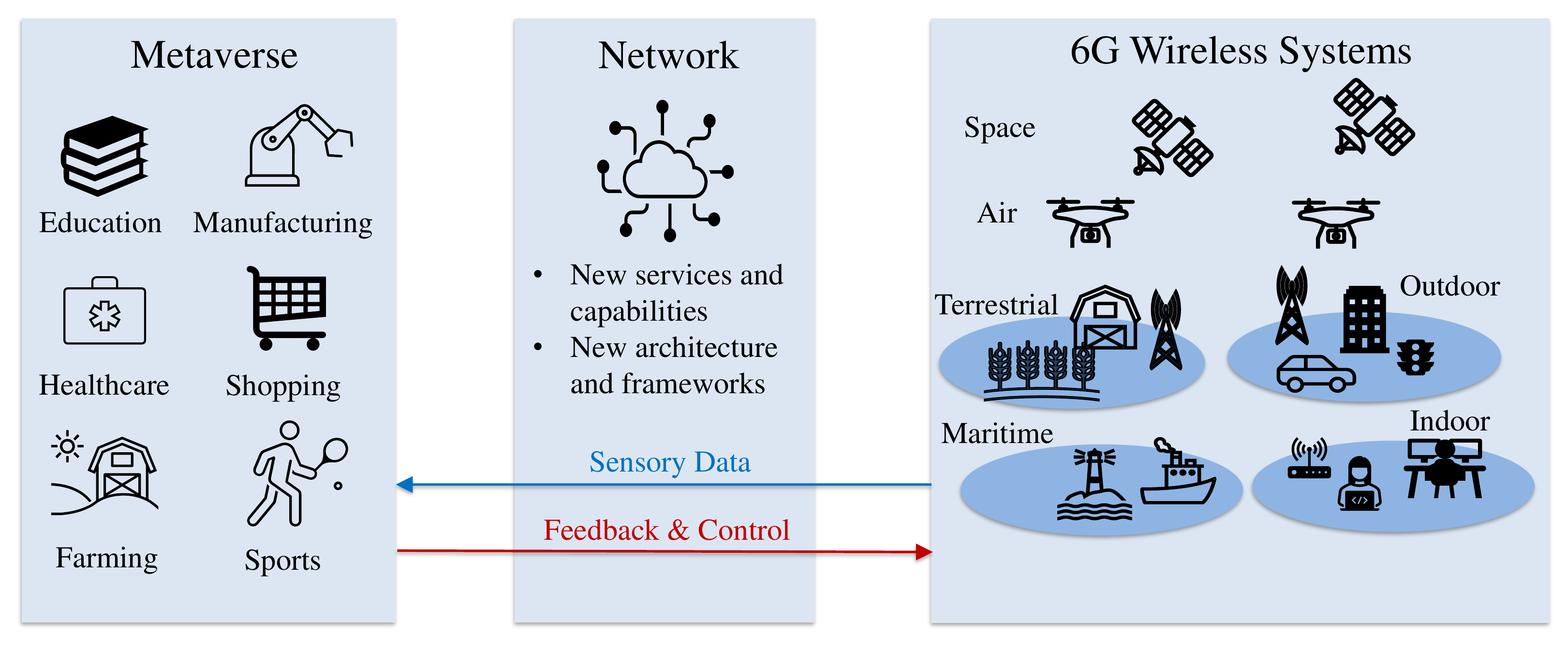}
			\caption{Illustration of the metaverse and 6G and beyond wireless systems.}
			\label{fig:sys}
		\end{figure}
		
		Generally, wireless systems for the metaverse mainly require the following two types of communications, as shown in Fig.~\ref{fig:gateway} and Fig.~\ref{fig:sys}.
		\begin{itemize}
			\item Human and Metaverse Communication (HMC). As shown at the top of Fig.~\ref{fig:gateway}, the communication between users and the metaverse involves specialized devices, including HTC displays, sensors, XR HMDs, legacy computers, and smart devices, that can be used to view and interact with the metaverse. These devices have sensing units that track users' motions and gather input data for computing in the metaverse. The computing results are sent back to users' devices for display and actuation.  
			\item Machine and Metaverse Communication (MMC). Active and passive sensors can monitor physical objects, machines, and environments. The aggregated data can be used to develop hyperreal virtual models in the metaverse. Additionally, the metaverse can use actuators to interact with and change the physical world to meet users' request. The communication between the metaverse and sensors and actuators can use 6G wireless systems that support IoE.
		\end{itemize}
		
		It is important to note that HMC and MMC are bidirectional, reflecting the interactions between the physical and virtual worlds. For human-to-human and machine-to-machine communications, these interactions can take place in the physical world without the need for the metaverse, or they can use the metaverse as a gateway to become HMC or MMC. Mulsemedia communication is utilized to improve the QoE, and it is used for HMC. 
		
		Currently, the most popular tool for accessing the metaverse is the XR HMD \cite{akyildiz2022wireless}. However, some users experience frustration, eye strain, and sickness from prolonged use due to the large weight and low resolution of HMDs \cite{kalamkar2022quantifying}. While working and living in the metaverse provides various choices and support that are not available in reality, existing HMDs are not convenient and can even create anxiety for some users, according to a recent study \cite{kalamkar2022quantifying}. The low-quality of metaverse content, constrained by communication, computing, networking, and sensing capabilities, is the primary cause. Moreover, the HMDs integrate many electronic modules that have non-negligible weights, such as the battery which determines the operation time of wireless HMDs.
		
		\subsection{6G wireless systems}
		
		Users of the metaverse expect hyperreal and immersive virtual content that are compute and data-intensive \cite{cai2022joint}. Interaction with other users or virtual content in the metaverse demands ultra-low end-to-end latency and ultra-high reliability to avoid causing discomfort, such as dizziness, to users. As a result, the metaverse requires ultra-high data rates, ultra-high reliability, and ultra-low latency communications with computation capabilities in both core and edge networks. While 5G wireless systems have started the creation and development of the metaverse, providing high QoE services for the metaverse remains a challenge. 5G wireless systems are equipped with a range of services that can support metaverse applications, especially the enhanced Mobile Broadband (eMBB) and Ultra-Reliable Low Latency Communication (URLLC). eMBB provides high data rates and the URLLC prioritizes high reliability and low latency. However, the metaverse demands technologies that can integrate both of them simultaneously. It also requires various other services that are not available in 5G wireless systems, such as integrated communication and sensing and the deterministic networking with bounded end-to-end latency \cite{saad2019vision,akyildiz2022holographic}.

		The 6G wireless system \cite{akyildiz20206g,saad2019vision} is anticipated to provide ubiquitous, high-speed, low-latency, and ultra-reliable connectivity simultaneously, which is the convergence of the eMBB, URLLC, and other services of 5G wireless systems. These innovative services will be made possible by advancements in all-spectrum communications, including mmWave communications, terahertz communications, optical communications, joint communication and sensing, space-terrestrial integrated networks, cubesat networks, deterministic networks, mobile edge computing, artificial intelligence, and other technologies, as shown in Fig.~\ref{fig:6G}. The physical world will be fully interconnected from space to air, terrestrial, maritime, indoor, and underground environments via 6G wireless systems, as shown in Fig.~\ref{fig:sys}. This will provide global coverage and seamless connection between the virtual world and the physical world. The cloud and edge networks, equipped with new architectures and frameworks, can facilitate the delivery of new services and capabilities to support metaverse applications. The QoE of the metaverse is expected to be significantly improved by leveraging 6G wireless technologies in communications, networking, caching, and computation, and using flexible and lightweight wireless devices.

		\begin{figure}[t]
			\centering
			\includegraphics[width=0.47\textwidth]{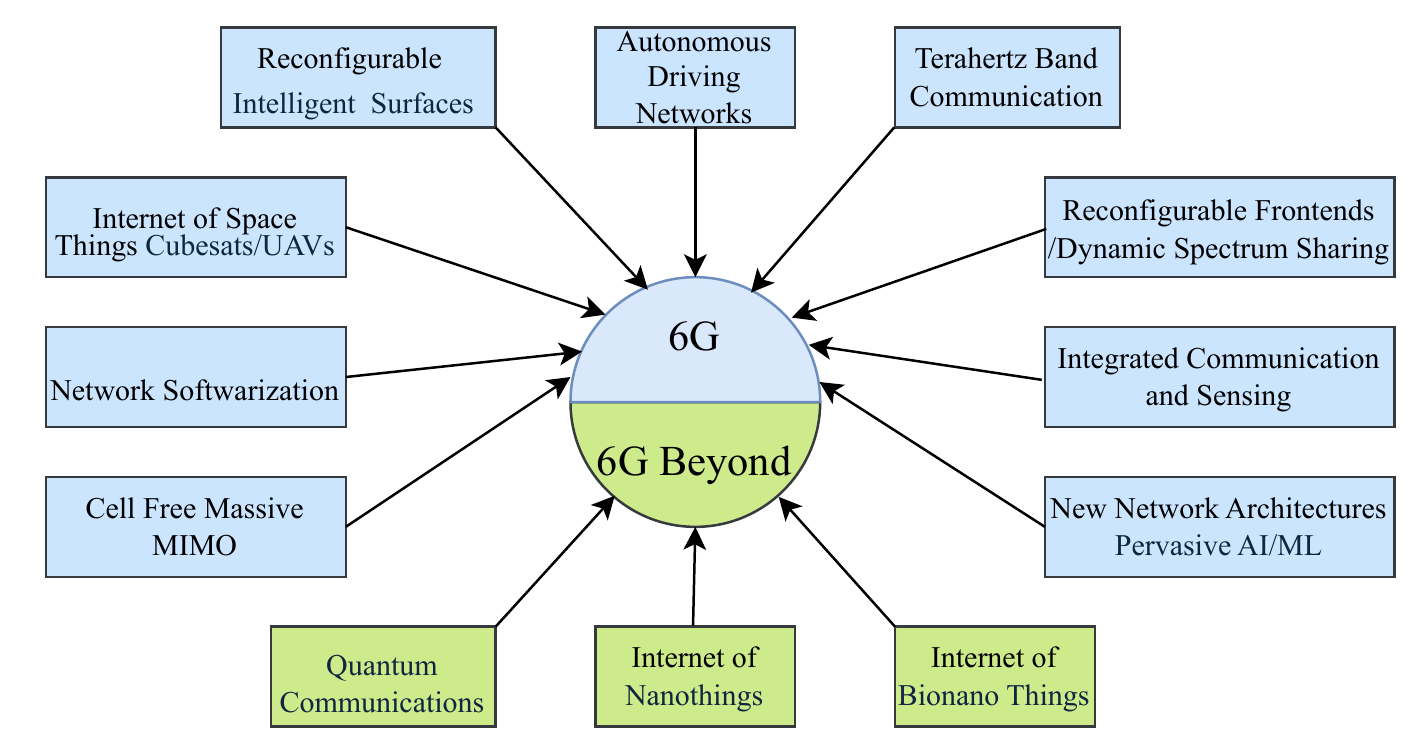}
			\caption{6G and beyond technologies \cite{akyildiz20206g}.}
			\label{fig:6G}
		\end{figure}

		\subsection{XR and HTC}
		
		XR is a type of spatial computing technology that enables users to observe and interact with virtual worlds \cite{shekhar2015spatial}. According to Milgram and Kishino's Reality-Virtuality Continuum \cite{milgram1995augmented}, XR technologies can be categorized based on their virtual content percentage and complexity. Augmented Reality (AR) only contains simple virtual content, with the majority of observations still being related to the real physical environment. Virtual Reality (VR) has full virtual content, with observations not related to the real physical environment. Mixed Reality (MR) combines both real and virtual content, with the virtual content often being interactive and dynamic. While there is overlap between MR and AR, MR is a broader concept that includes more interactive and richer virtual content. 
		
		Mulsemedia has been recently introduced to XR to provide more immersive user experiences \cite{OVR}. XR devices can display and output various mulsemedia, including 360-degree videos, haptics, and scents. While smart devices like smartphones and tablets can be used to access XR content, they cannot provide highly immersive experiences as users must look at the screen. High-end AR and MR devices use optical see-through head-mounted displays (HMDs) to seamlessly integrate virtual content with the physical environment. Conversely, VR uses HMDs that completely block the users’ view of the physical environment, and the displayed content is fully virtual. XR HMDs integrate various sensing, computing, communication, storage, and power units, including head tracking, hand tracking, eye tracking, proximity, and pressure sensors. These sensors enable XR devices to understand users’ behavior and input. Mulsemedia sensors and displays can be embedded in XR HMDs without significant modifications. However, XR HMDs have limited computing, storage, and energy due to size and weight constraints. As a result, they usually need to pair with a server, which can be a cell phone, tablet, computer, or remote edge/cloud server, to offload computation tasks and obtain extra storage.

		HTC is characterized by the delivery of holograms and other mulsemedia through wireless and wired networks \cite{akyildiz2022holographic}. A hologram is a recording of light field that preserves the original depth and parallax of real 3D objects. Hologram videos could provide 6 Degrees of Freedom (DoF) immersive viewing experiences, and users can freely move forward/backward (surging), up/down (heaving), or left/right (swaying) to select their favorite viewing angle of a 3D scene \cite{liu2021point}. This advances the user experience over common VR applications, as VR video users are unable to move forward or backward freely. Furthermore, HTC systems may collect mulsemedia data using various sensors at the source and then regenerate the environment using various actuators, displays, and speakers at the destination. Users can select the senses depending on the available hardware and privacy and security issues. 
		
		{{A generic HTC system consists of the source, the destination, and the network \cite{akyildiz2022holographic}.} The source uses various sensors to capture holographic content, synchronizes multisensory data, encodes the holographic data, and follows HTC networking protocols to send data packets. The HTC networks deliver source data with guaranteed performance in terms of bandwidth, latency, reliability, etc, which are defined by the HTC-enabled use cases. The destination receives and renders data for display, utilizes various actuators to regenerate the environment at the source, performs synchronization among multisensory media, and provides useful feedback to the source and the network if necessary. Users can observe truly 3D holograms with naked eyes using light field displays. Together with other mulsemedia content, HTC can provide a high quality of experience. {{Typical HTC use cases can be found in \cite{akyildiz2022holographic}.}
				
				\begin{figure}[t]
					\centering
					\includegraphics[width=0.47\textwidth]{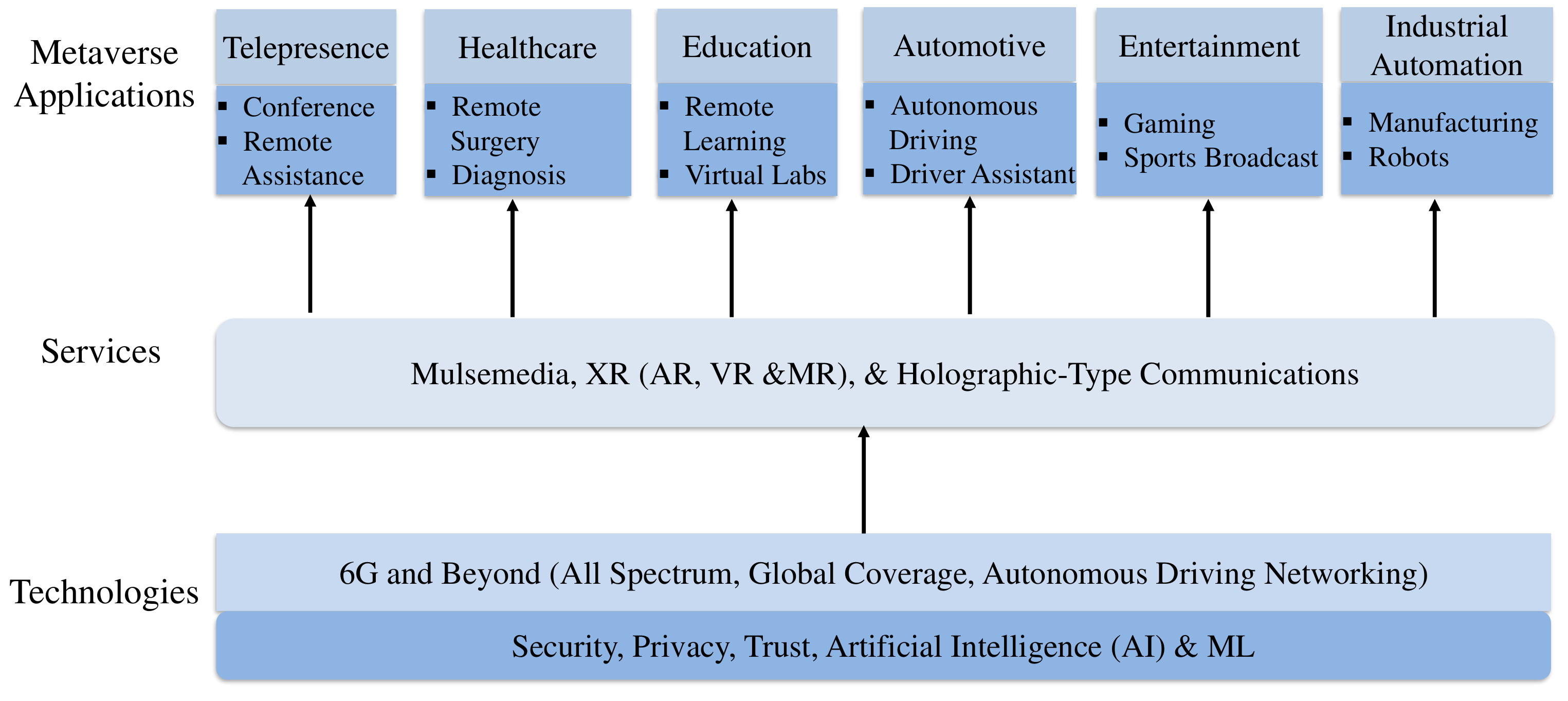}
					\caption{Metaverse applications, accessing technologies, and enabling technologies.}
					\label{fig:metaverse}
				\end{figure}
				
				\begin{figure}[t]
					\centering
					\includegraphics[width=0.37\textwidth]{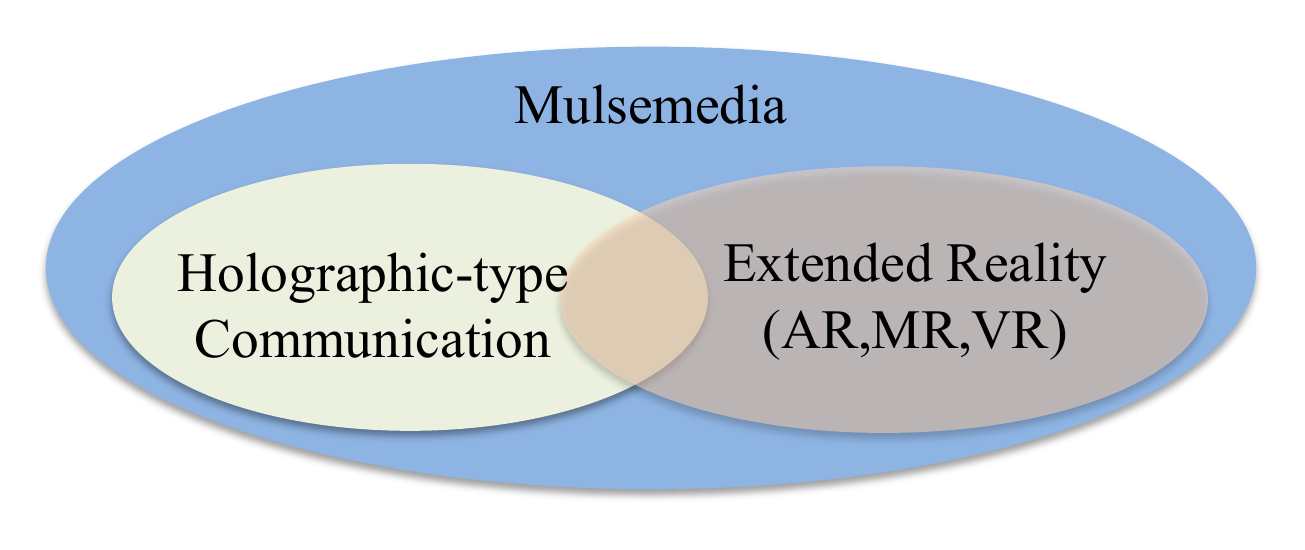}
					\caption{Metaverse and immersive mulsemedia technologies.}
					\label{fig:meta_world}
				\end{figure}
				The roles of mulsemedia communication and the associated XR and HTC technologies of the metaverse are shown in Fig.~\ref{fig:metaverse}. The three key aspects of the metaverse are its applications, services (mulsemedia, XR, and HTC), and supporting technologies, especially the 6G wireless systems, as shown in Fig.~\ref{fig:6G} (more details of the 6G and beyond technologies can be found in \cite{akyildiz20206g}). Metaverse applications and users create the virtual worlds with social and economic activities, and XR and HTC are the key accessing technologies used to enter these virtual worlds. The relation between the metaverse, mulsemedia, XR, and HTC is shown in Fig.~\ref{fig:meta_world}. Mulsemedia communication is a broad concept that includes all media that stimulate multiple human senses. XR and HTC are technologies that can deliver mulsemedia content. Mulsemedia can also be recorded and displayed using other traditional sensors and displays with reduced QoE. There are many ways to display holograms where XR HMD is one of these. XR HMDs can show different views, which can create 3D effects to display holograms. As a result, there is an intersection between XR and HTC. 
				
				\section{Use Cases}
				In this paper, we explore four crucial use cases for mulsemedia communication in the fields of education and training, healthcare, retail, and entertainment. These domains are essential in our daily lives and provide an excellent opportunity to develop new mulsemedia communication theories, technologies, and business models. {{The selected use cases are representative of a broad spectrum of applications and demonstrate the varied potential applications of mulsemedia communication; however, they do not encompass the full scope of possible scenarios. }}

				\subsection{\bf Education and training} 
				Combining video and audio with sensory signals such as scent, haptic feedback, light, temperature, and more can create immersive learning experiences. For instance, students can touch, smell, and see flowers in biology and art classes using mulsemedia. This multisensory approach can enhance students' understanding and promote their learning experiences in various subjects, such as geology, where they can experience different weather patterns and conditions worldwide. Furthermore, mulsemedia can be a valuable tool for training professionals, such as first responders, miners, and farmers, as it enables them to recognize and fully grasp critical incidents and respond effectively. Mulsemedia can simulate the scents, temperature changes, and flashlights that professionals may encounter, making training more realistic and engaging.  
				
				\subsection{\bf Healthcare} 
				Telehealth technology uses computer networks to connect healthcare providers with patients, improving healthcare accessibility and enabling doctors to treat contagious diseases without being in the same location as their patients. While traditional multimedia can only help to provide a basic diagnosis, patients must answer many related questions and describe their feelings. Mulsemedia communication can significantly increase diagnostic accuracy by allowing doctors to leverage more sensory information, such as haptic and temperature signals. Additionally, remote surgeries can be performed by doctors using mulsemedia communications to control robots or robotic arms/hands using haptic signals.
				
				\subsection{\bf Retail} 
				The retail industry has traditionally relied on strategies such as sending samples or using multimedia-based advertisements. However, customers often have to return goods because they lack sufficient information, which can only be obtained after receiving them. Mulsemedia offers a solution by providing a comprehensive description of a product through various senses. By using mulsemedia-based descriptions, customers can accurately experience the material, smell, and condition of goods before making an online purchase. As a result, this can increase customer satisfaction and provide greater success and revenues for retailers.
				
				\subsection{\bf Entertainment and social interactions} 
				{{Mulsemedia technology can be integrated into various existing smart devices for entertainment and social interactions. Mulsemedia can be used to develop truly immersive entertainment environments by stimulating more human senses. In 2011, Samsung collaborated with the University of California at San Diego to develop a smart TV that can generate over 10,000 different odors. Additionally, smartphones, tablets, smartwatches, and other smart devices can process and display mulsemedia and allow users to interact with each other using various sensory media. This technology can be used to augment existing videos and pictures with additional senses, providing more immersive experiences. For instance, instead of sending simple videos, audios, or text messages, we can send friends various sensory messages, e.g., scent of followers.}}

				\section{Mulsemedia Communication Systems} 
				
				Existing multimedia communication schemes can be broadly divided into two categories, namely, Video-on-Demand (VoD) and Live Streaming (LS). The VoD content is prerecorded and streamed to users based on requests, while LS content is recorded and immediately sent to users. LS is more challenging than VoD since only limited time is available to record and preprocess multimedia content. Similarly, mulsemedia communications comprise of Senses-on-Demand (SoD) and LS, where ``video'' is replaced by ``senses''.   
				
				Existing mulsemedia communication systems are mainly SoD; LS remains a challenge. Although they require different streaming technologies, SoD and LS share the fundamental system units. Next, we introduce the three major components of a mulsemedia communication system, namely, the sources, the networks, and the destinations, as shown in Fig.~\ref{fig:streaming}.

				\begin{figure*}[t]
					\centering
					\includegraphics[width=0.7\textwidth]{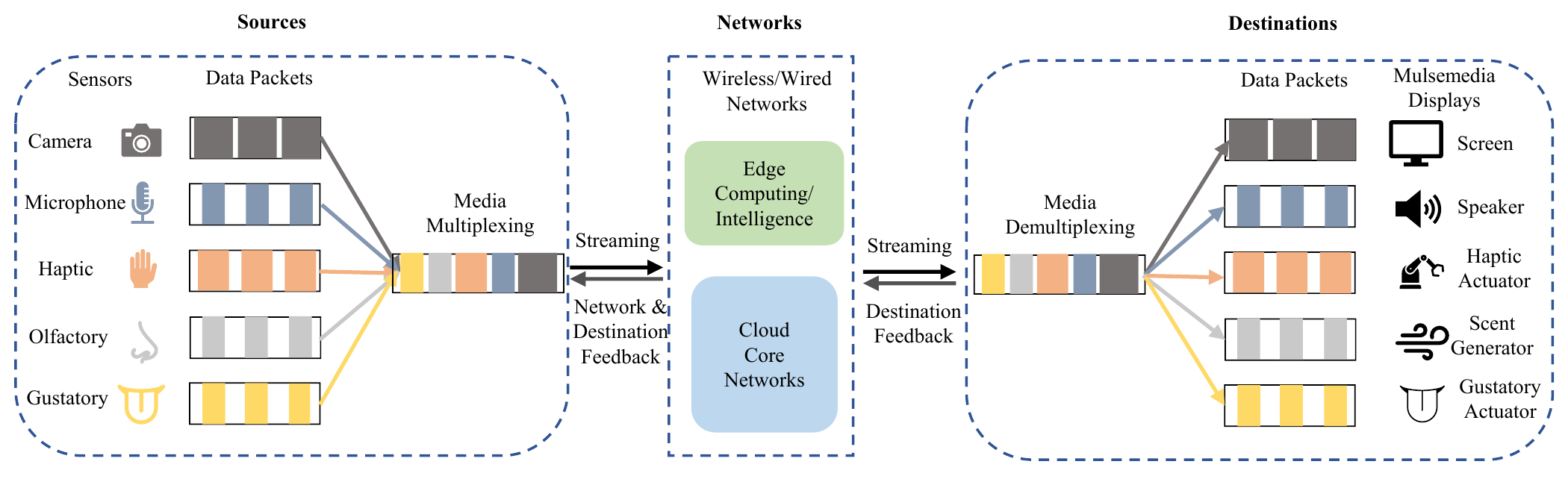}
					\vspace{-5pt}
					\caption{Mulsemedia communication system with sources, networks, and destinations. Multisensory content is multiplexed at the sources and demultiplexed at the destinations. }
					\label{fig:streaming}
					\vspace{-0pt}
				\end{figure*}
				\subsection{Source}  
				The source of SoD can be a server where mulsemedia is stored, while the source of LS can come from various devices such as computers, smart devices, sensory booths, or XR headsets. There are two methods of creating mulsemedia:
				\begin{itemize}
					\item \emph{Annotation}. Additional senses can be added to existing multimedia files using annotation tools, such as PlaySEM \cite{saleme2015playsem}. Authors typically perform the annotation based on personal judgment. For instance, when a video shows a flower, the author can annotate a smell based on previous experiences or knowledge, which may not necessarily be the exact scent of the flower.
					
					\item \emph{Real-time recording}. Multiple senses are captured using various sensors simultaneously, and the sensing data is processed and encoded to create mulsemedia. The quality of mulsemedia in this approach is dependent on the sampling rate and accuracy of the sensors. However, this method results in a larger data size and higher data rates for communication compared to the annotation approach.  
				\end{itemize}
				
				As mentioned earlier, the annotation-based approach for creating mulsemedia is relatively straightforward, and existing standards like MPEG-V \cite{waltl2013end,yoon2015mpeg}, which will be discussed in Section \ref{sec:standards}, have already provided comprehensive solutions. Therefore, we will focus on the real-time recording-based approach in the following sections.
				
				\subsubsection{Sensors} 
				While mulsemedia communication poses several challenges, off-the-shelf sensors can capture various senses \cite{sulema2016mulsemedia}. These include cameras, microphones, light intensity sensors, moisture sensors, temperature sensors, and wind sensors. Emerging haptic, olfactory (related to smell), and gustatory (related to taste) sensors are also available, as shown in Fig.~\ref{fig:sensor}. In this context, all displays and actuators used to replay mulsemedia are called displays. These sensors and displays can be integrated with existing smart devices, computers, and XR HMDs, as shown in Fig.~\ref{fig:sensor1}. 
				
				Aside from the five basic senses, human beings also have many other senses, such as skin senses of moisture, temperature, and pressure. It is challenging to integrate all available sensors in XR HMDs due to the space and weight constraints. It is anticipated that XR HMDs need to be used together with wearable sensors to offload its complexity, weight, and power consumption. Furthermore, the HTC user with light field displays do not need any wearable devices to obtain a high QoE. Thus, novel designs of sensors without direct contact with the user's body are required. 
				
				Mulsemedia recording utilizes all available sensors to capture synchronized sensing media. While different sensors may have unique characteristics, the fundamental sensing mechanisms are similar. Sensors sample the sensing input signals periodically and compress and encode the resulting data. However, since sensing dynamics vary, for example, the environment temperature changes much slower than human motion, mulsemedia sensors require different sampling rates, data throughput, and sensitivity that depend on specific applications. 
				
				\begin{figure}[t]
					\centering
					\includegraphics[width=0.48\textwidth]{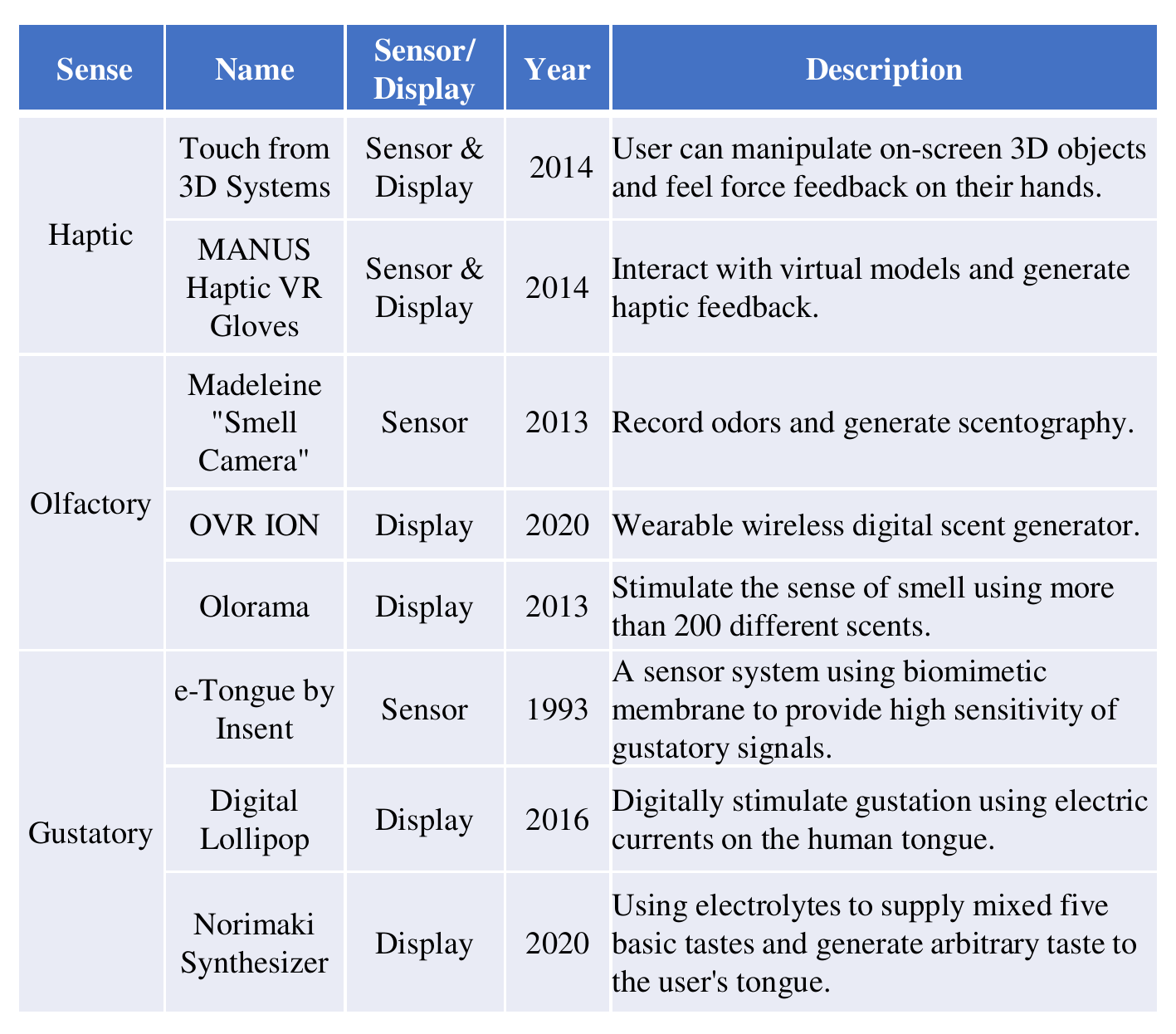}
					\vspace{-5pt}
					\caption{Mulsemedia sensors and displays. The sensors and displays are 3D systems \cite{3dsystems}, MANUS Prime X Haptic VR gloves \cite{MANUS}, Madeleine Smell Camera \cite{madeleine}, OVR ION \cite{OVR}, Olorama \cite{Olorama}, e-Tongue \cite{insent2023}, Digital Lollipop \cite{ranasinghe2016digital},  and Norimaki Synthesizer\cite{miyashita2020norimaki}.   }
					\label{fig:sensor}
					\vspace{-0pt}
				\end{figure}
				
				\begin{figure}[t]
					\centering
					\includegraphics[width=0.48\textwidth]{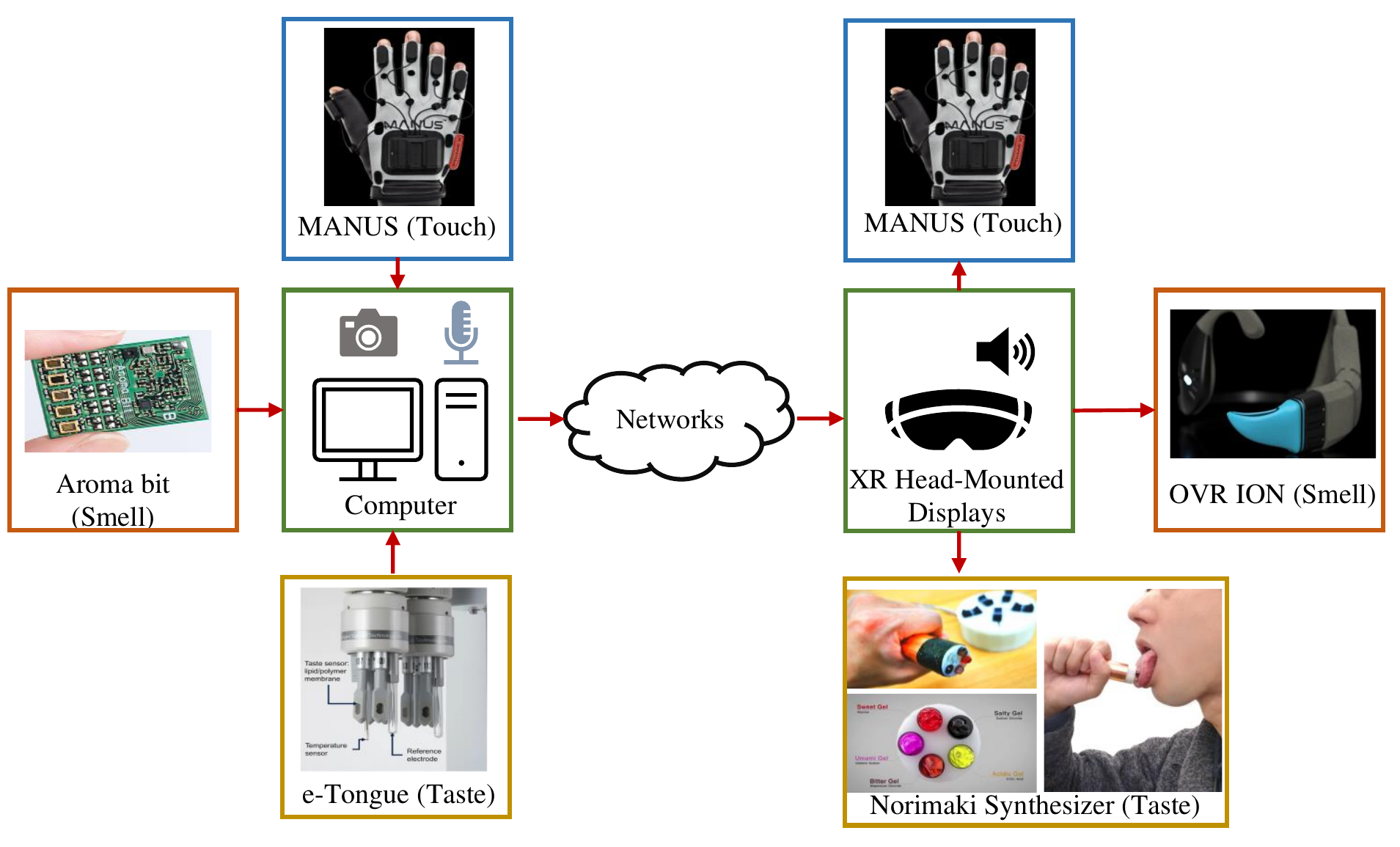}
					\vspace{-5pt}
					\caption{Mulsemedia communication system with a source (computer) and a destination (XR head-mounted display). The sensors and displays are e-Tongue \cite{insent2023}, Aroma bit \cite{aromabit}, MANUS Prime X Haptic VR gloves \cite{MANUS}, OVR ION \cite{OVR}, and Norimaki Synthesizer\cite{miyashita2020norimaki}.   }
					\label{fig:sensor1}
					\vspace{-0pt}
				\end{figure}
				
				\subsubsection{Mulsemedia encoding} 
				Mulsemedia encoding can be performed by first encoding each sensory media and then multiplexing them for streaming, as illustrated in the left part of Fig.~\ref{fig:streaming}. In this way, mulsemedia can be created by slightly modifying existing multimedia encoding methods, rather than designing new encoding systems from scratch. Furthermore, by increasing the sensing dimension and adding more sensory media to existing encoding systems, we can further enhance mulsemedia capabilities. Haptic, olfactory, and gustatory media are relatively new, and the corresponding encoding principles are similar to those used for other sensory media. {{For haptic signals, Apple uses AHAP format which is a JSON-like file. The IEEE 1918.1 Tactile Internet standard proposes haptic codecs including a kinesthetic codec and a tactile codec \cite{holland2019ieee}. There are commercially available olfactory sensors and displays, but dedicated international encoding standards for olfactory media is currently not available. Gustatory media encoding faces a similar issue.}}   
				
				Fig.~\ref{fig:tsn} shows the required data rates for each sensory media. As we can see, HTC requires the highest data rates due to the high complexity of holograms. Thus, besides inter-media encoding, the major intra-media encoding challenge lies in HTC. The current trend in presenting holograms is to use point cloud videos. A point cloud is a set of 3D points, where each point is associated with a geometry position and attribute information \cite{graziosi2020overview}, such as color, reflectance, transparency, etc. 3D point clouds could be generated by synthesizing the videos captured simultaneously by multiple RGB cameras at different angles or multiple RGB cameras and depth cameras. For example, in the 8i Voxelized Surface Light Field (8iVSLF) dataset \cite{krivokuca20188i}, the full body of a human subject was captured by 39 synchronized RGB cameras configured in either 12 or 13 rigs (or clusters), at a frame rate of 30 fps. The camera rigs were placed around the subject at approximately a few meters distance. Each cluster of cameras captured RGB and computed depth-from-stereo. The inputs from all data clusters were then fused into a 3D surface.
				
				HTC systems usually include compression solutions to reduce the bandwidth demand of raw point cloud videos. MPEG Video-based Point Cloud Compression  (MPEG V-PCC) and MPEG Geometry-based Point Cloud Compression (MPEG G-PCC) are the major standards for point cloud compression \cite{graziosi2020overview}. MPEG V-PCC encodes 3D point clouds as a set of 2D videos leveraging existing 2D video compression standards and tools. It is well suited for dense sequences that, when projected to 2D, result in image-like continuous and smooth surfaces. MPEG G-PCC encodes the content directly in 3D space. As the current G-PCC standard has only defined intra-prediction but no temporal prediction, it is more suitable for sparse meshes and static scenes. 
				
				Recent encoding experiments with MPEG V-PCC have demonstrated its good compression performance, producing decent quality for test point cloud sequences with a bit rate of 10 - 25 Mbps \cite{graziosi2020overview}. However, the encoding process for point cloud videos is time-consuming. As reported in \cite{liu2019comprehensive}, for a set of point cloud test sequences over a 7-10s period, the average coding time using the MPEG TMC13 software took as long as 8.28 seconds. There is a need to reduce the processing latency of encoding for highly interactive and live broadcasting use cases. At the destination side, billions of devices are already capable of decoding V-PCC content, based on the readily available hardware 2D video decoding solutions. However, it is still challenging to implement V-PCC decoders on mobile devices, as decoding a V-PCC bitstream requires the synchronization of three video decoder instances, which is not adequately supported by current mobile operating systems \cite{schwarz2019real}. The decoder also performs additional computations to render the 3D content for display. Therefore, the latency of the decoding process should also be optimized for live streaming scenarios.
				
				In addition to the MPG V-PCC and G-PCC compression standards, there are a few other studies that address the compression of 3D/point cloud. Draco \cite{google} is a software library (by Google) that aims to compress 3D geometric meshes and point clouds to enhance the storage and transmission of 3D graphics. Draco continuously splits the point cloud from the center utilizing the concept of KD tree formation, while also modifying the axes on each direction, followed by the application of entropy encoding techniques to compress the data \cite{bui2021comparative}. Deep learning-based point cloud compression methods utilize neural networks to learn the underlying patterns and structures in the point cloud data, enabling them to efficiently represent the data with fewer bits. These methods typically use autoencoder networks or convolutional networks to learn a compact representation of the input data, which can then be decoded to reconstruct the original point cloud \cite{huang20193d,quach2020improved,wang2021lossy}. These methods have shown promising results in terms of compression rate and reconstruction quality, making them a potential solution for efficient storage and transmission of 3D point cloud data. However, these methods also have their own challenges, such as the need for large amounts of training data and the computational complexity of the network architectures.

				\subsection{Communication and networking} 
				
				\begin{figure}[t]
					\centering
					\includegraphics[width=0.45\textwidth]{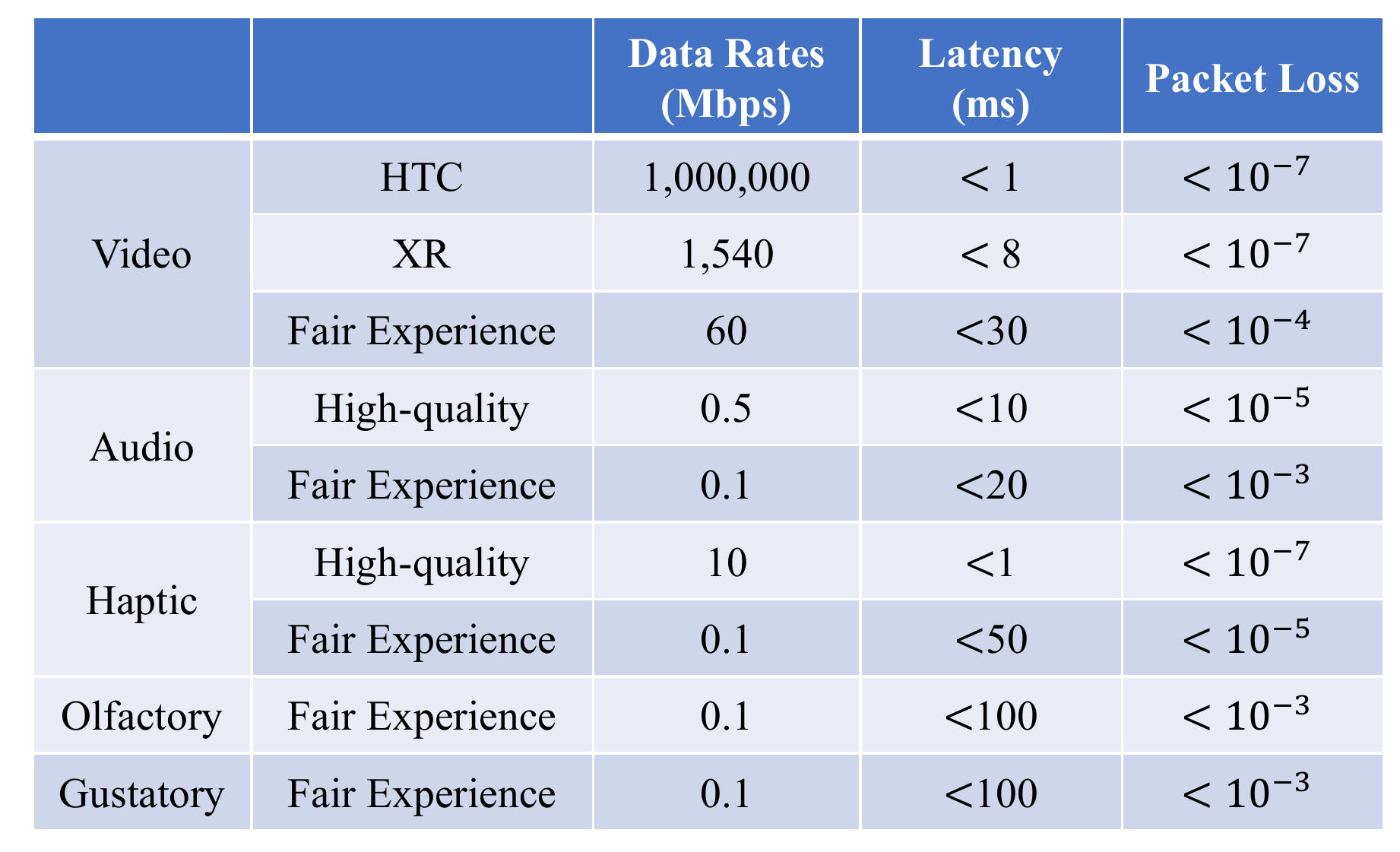}
					\vspace{-5pt}
					\caption{Estimated required data rates, end-to-end latency, and packet loss rates for five basic sensory media \cite{holland2019ieee,ITU}.}
					\label{fig:tsn}
					\vspace{-0pt}
				\end{figure}
				
				The annotation-based mulsemedia can be streamed through existing wireless and wired networks since it only adds negligible annotation metadata. Existing multimedia protocols can be modified to add more sensory media for mulsemedia streaming, which will be discussed in Section \ref{sec:standards}. The research efforts that aim to improve data rates and reduce end-to-end latency in Wi-Fi and 5G wireless systems can be leveraged for mulsemedia communication. 
				
				Over-The-Top (OTT) multimedia streaming services could be used to adapt the transmission of mulsemedia  according to network conditions. Current OTT streaming solutions are based on either the Transmission Control Protocol (TCP) or User Datagram Protocol (UDP). Retransmission is used in TCP when a packet is lost or not successfully received, which will increase latency and bandwidth requirements. Therefore, TCP-based solutions are more suitable for low-interactive applications such as the streaming of prerecorded content. A popular TCP-based solution for streaming prerecorded videos is Dynamic Adaptive Streaming over HTTP (DASH). DASH-based solutions for point cloud video streaming have been proposed recently \cite{liu2021point}. At the server side, point cloud videos are divided into tiles and encoded under different quality levels, which can be selectively transmitted to save bandwidth. A tile selection module at the client side calculates the tiles residing in the user’s Field-of-View (FoV) and selects the proper quality level for each tile to maximize the viewing experience. A DASH-based mulsemedia communication testbed has been developed in \cite{bi2018dash}, where the synchronization issue is addressed and users obtain a high level of satisfaction. 
				
				UDP-based protocols are in favor of lower latency at the expense of a decreased reliability, and therefore, they have the potential to support high-interactive or LS mulsemedia applications. Traditional UDP-based streaming solutions, such as Web Real-Time Communications (WebRTC), work well for streaming 2D videos, but they need more investigation and redesign to render them suitable for transmission of mulsemedia data. Furthermore, to meet users' reliability requirements, UDP-based protocols should be combined with loss recovery technologies such as error-resilient encoding at the source and error concealment at the destination.
				
				The transmission of LS mulsemedia poses significant challenges due to the high volume of sensory data flows from various sensors such as haptic, olfactory, gustatory, and others. These data flows can be multiplexed at the source and demultiplexed at the destination to reduce the complexity of multisensory data synchronization, as shown in Fig.~\ref{fig:streaming}. However, each sensory data flow has its unique characteristics, and it is more efficient to consider the embedded data structure and spatial-temporal correlations when encoding the data. Some sensors can record the data and directly send it through the networks to corresponding displays at the destination, reducing the end-to-end latency by simplifying the encoding and decoding processes. However, this method also has its challenges. First, the synchronization of multiple data flows can be difficult due to the variance in the end-to-end latency of the best effort delivery service. Second, it requires large buffers and coordination of the displays, leading to large end-to-end latency. In general, both architectures have limitations and advantages. A hybrid system that can support both multiplexed and independent multisensory data flows may find a balance.

				\subsection{Destination} 
				
				The destination of mulsemedia communication plays a critical role in the overall system. It performs four main tasks to ensure the successful reproduction of mulsemedia content. First, it receives and decodes the mulsemedia data packets that are transmitted over the network. Decoding can introduce some latency depending on the compression ratio used during encoding. Additionally, in case of transmission errors, the destination may request retransmission or attempt to correct or conceal the errors. Second, it synchronizes the multisensory data that arrive at different times. The destination must compensate for any variations in latency that may arise due to network delays or other factors. Third, the destination needs to coordinate the recreation of the mulsemedia using various displays, such as actuators, speakers, monitors, etc. This task involves optimizing the scheduling of display activities considering the unique characteristics of different devices. Finally, the destination can provide feedback to the source regarding the status of the network for adaptive transmission. By performing these tasks, the destination ensures that mulsemedia content is accurately and efficiently reproduced, leading to a high-quality user experience.

				Due to the high volume of hologram data, recreating the hologram is the most complex task at the destination. There are two major types of displays that allow the viewing of holograms with naked eyes with a much better user experience than by using XR HMDs or smart devices: multiview volumetric display and light field display. The multiview volumetric display is a relatively small holographic display that can be used for mobile devices and holographic monitors. Usually, it can only support one user, and eye-tracking technologies can be used to adaptively render the holographic content. Light field displays can provide thousands of view angles and support multiple users simultaneously. Light field displays are anticipated to be widely used in future HTC and mulsemedia systems since they can provide a hyperreal user experience. To display real-time holograms of human subjects using a 1.5 m $\times$ 0.75 m display with 30 fps, a light field display requires an uncompressed data rate of 6.63 TBps \cite{burnett201761}. Such a high data rate is beyond the transmission capacity of today's networks, which necessitates new designs of wireless communication systems to support mulsemedia communications with light field displays at the destination.  
				
				\subsection{Standards}
				\label{sec:standards}
				MPEG-V is an international standard for mulsemedia communication which consists of the following parts: architecture, control information, sensory information, virtual world object characteristics, data formats for interaction, common types and tools, and reference software and conformance \cite{yoon2015mpeg}. Its main goal is to enhance interactions between the real and virtual worlds and support XR applications using mulsemedia. The sensory information part of the standard defines various senses, including light, temperature, wind, vibration, sprayer, scent, fog, color correction, motion, kinesthetic, and tactile. This sensory information is annotated as Sensory Effect Metadata (SEM) and streamed together with multimedia files as shown in Fig.~\ref{fig:xml_example}. The SEM file provides detailed information about the multisensory media, including the sense type, intensity, duration, start time, and other attributes. The lower part of Fig.~\ref{fig:xml_example} shows an example of the sensory effect description in XML. At the destination, the SEM file is decoded and synchronized with the multimedia file.
				
				The SEM created using MPEG-V has a negligible size which can be easily attached to existing multimedia formats. However, the multisensory effects defined in SEM are simple and only based on annotated descriptions. On the contrary, videos and audios are created using sampling which produces many samples and data packets in one second. Despite MPEG-V providing multisensory effects, it is challenging to achieve high QoE mulsemedia and realize LS of mulsemedia.   
				
				The inclusion of more sensory information in XR content is a growing trend in the multimedia communication field. The MPEG Immersive Video (MIV) standardization effort is an example of this, as it aims to incorporate haptic effects in videos with 6 DoF, providing users with a more immersive experience in the metaverse and other XR applications. Despite significant advancements in video and audio technologies, the development of other sensory media has been lagging behind. This presents a challenge for mulsemedia communication researchers to overcome. In the next section, we will explore some current research challenges and future opportunities in this field.   
				
				\begin{figure}[t]
					\centering
					\includegraphics[width=0.48\textwidth]{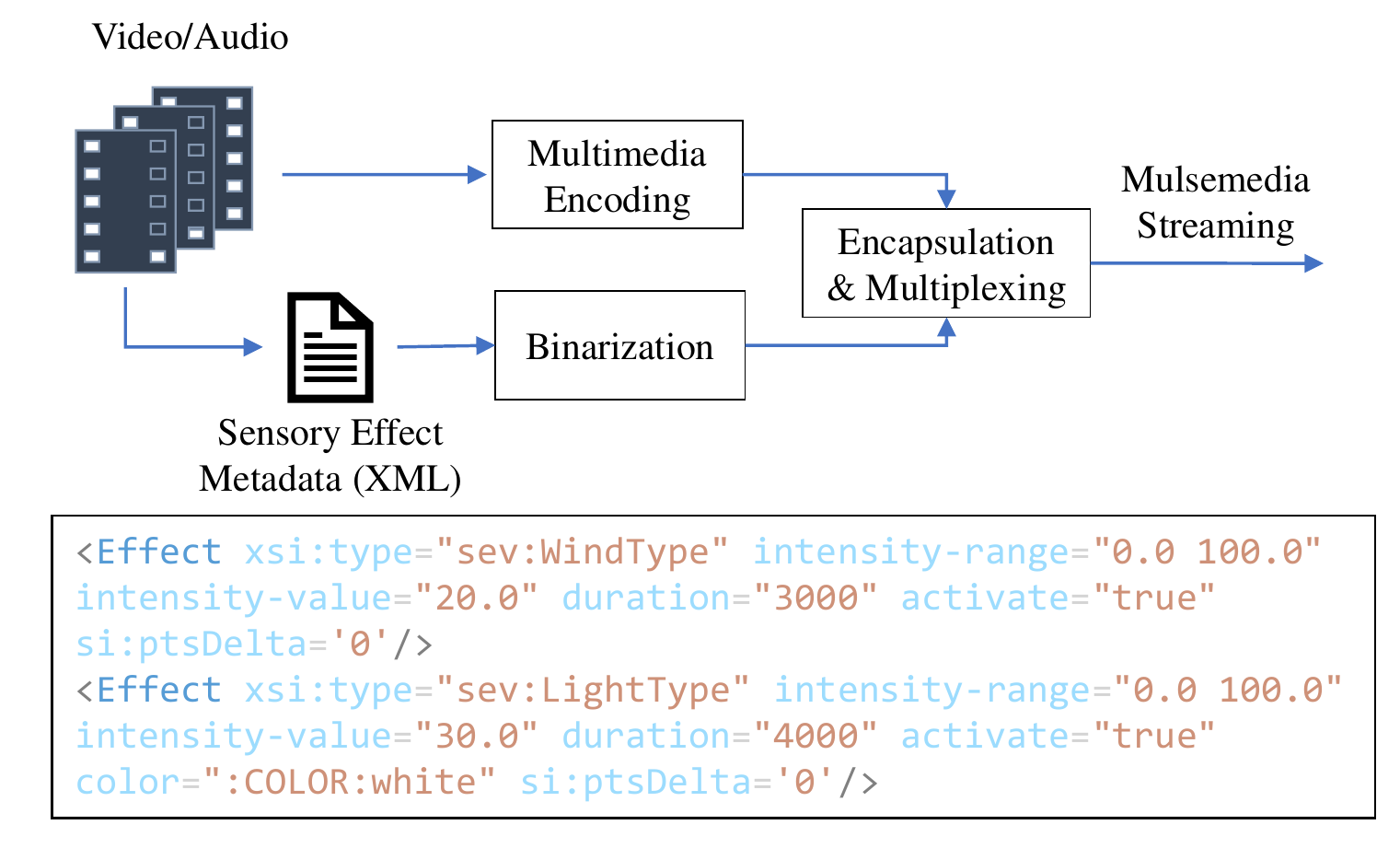}
					\vspace{-5pt}
					\caption{Mulsemedia transmission using MPEG-V. The lower part is an example of the sensory effect metadata which describes the effects of wind and light.}
					\label{fig:xml_example}
					\vspace{-0pt}
				\end{figure}
				
				{{The IEEE 1918.1 Tactile Internet standard focuses on haptic signal transmission through the Internet \cite{holland2019ieee}. The haptic signal can be sent together with videos and audios and thus the Tactile Internet can be considered as another mulsemedia application. The haptic data size depends on the DoF. For immersive virtual reality, the DoF ranges from 1 to 6, while for complex tele-operation the DoF can be as high as 100. The haptic codec consists of kinesthetic codec and tactile codec. The kinesthetic codec is developed for 3D position, velocity, force, and torque data, which considers two different scenarios, namely delay intolerant and delay tolerant. The tactile codec considers hardness, thermal conductivity, friction, microroughness, and macroroughness with relaxed delay requirements. }}

				\begin{figure}[t]
					\centering
					\includegraphics[width=0.49\textwidth]{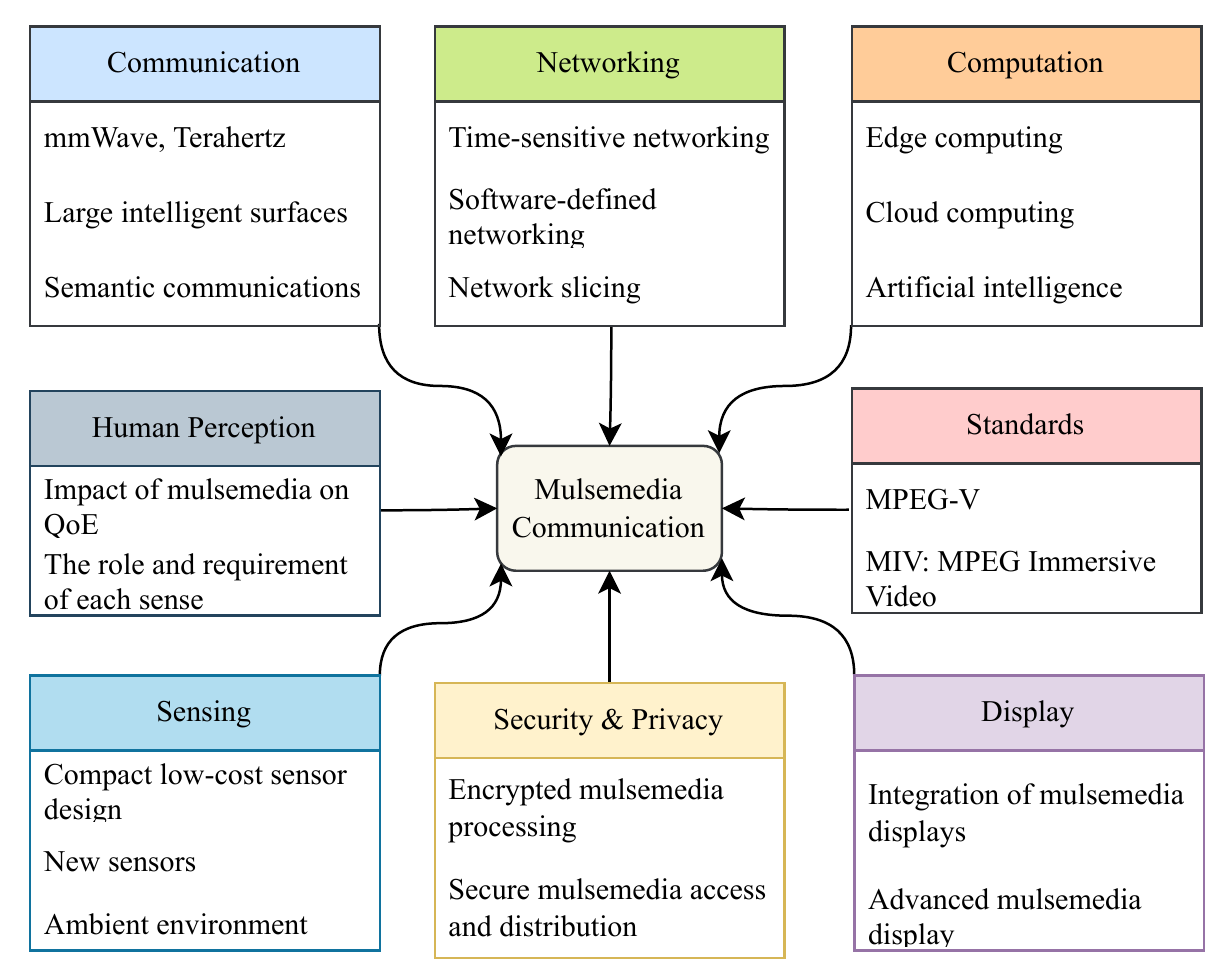}
					\vspace{-5pt}
					\caption{An overview of mulsemedia communication research areas and technologies.}
					\label{fig:arch}
					\vspace{-0pt}
				\end{figure}
				
				\section{Research Challenges and Potential Solutions}
				\label{sec:6G}
				This section delves into the research challenges and potential solutions for mulsemedia communication. An overview of the research challenges and areas is given in Fig.~\ref{fig:arch}. We consider XR and HTC are the technologies used for mulsemedia communications. The following technical requirements are essential for mulsemedia communications:
				\begin{itemize}
					\item Reliable high-speed, low-latency wireless communication to enable unrestricted movement. Existing mulsemedia communication systems use sensors and displays networked by cables, which significantly limit users' mobility. Some existing XR HMDs are also tethered by cables. 
					\item Accurate high-fidelity wireless sensing technologies that can monitor and predict users' motion and render content adaptively. Integrated head, hand, and eye tracking sensors in existing XR HMDs are used, but the ultimate goal is to remove HMDs to allow users to play mulsemedia without wearing any devices. Wireless sensing of the motion of head, hand, body, and eyes can effectively free users.
					\item New Internet protocols and wireless networking technologies are needed to provide new services and capabilities for delivering mulsemedia content with different quality-of-service.
					\item Computing and content caching must move from the core cloud towards the network edge to reduce end-to-end latency and support highly interactive mulsemedia applications. Real-time solutions will require the consideration of processing tasks for mulsemedia content generation, compression, and rendering, which could be time-consuming.
					\item Computing and communication solutions have to be designed to boost user experience and leverage the characteristics of QoE for system optimization.
				\end{itemize} 
				
				5G wireless systems are unable to meet the requirements due to the following shortcomings. 
				\begin{itemize}
					\item {\bf Data rates}. While 5G wireless systems promise data rates exceeding 1 Gbps, recent measurements have shown that the throughput can be as low as 100 Mbps, which is insufficient for mulsemedia applications \cite{xu2020understanding,rischke20215g}. Although mmWave can provide a higher throughput of around 2 Gbps, the throughput is not stable and requires more intelligent management to maintain the desired performance \cite{narayanan2020lumos5g}. As shown in Table~\ref{tab:comparison2}, the experienced data rates are much smaller than the peak data rates. 
					\item {\bf End-to-end latency}. 5G wireless systems promise an end-to-end latency of 1 ms, but recent measurements have shown that the latency is much higher due to constraints in the wire-line networks \cite{xu2020understanding}. Private networks with reserved resources can reduce the latency significantly, but it still remains higher than 1 ms \cite{rischke20215g}.  
					\item {\bf Reliability}. 5G wireless systems achieve much higher reliability than 4G wireless systems \cite{rischke20215g}. For 5G standalone networks, the core network packet loss rate is on the order of $10^{-6}$\% and the radio access network packet loss rate is 3.28$\cdot 10^{-4}$\%, as measured in \cite{rischke20215g}. However, for large size packets, which are common in mulsemedia applications, the packet loss rate increases significantly at the air interface to around 8.17\% \cite{rischke20215g}.  
				\end{itemize}  
				
				In addition to these limitations, 5G wireless systems have shortcomings with respect to energy efficiency, coverage, sensing, computing, and caching. On the contrary, 6G wireless systems have the potential to address most of these challenges. In the following, we will discuss the challenges to meeting mulsemedia communication requirements and their potential solutions.

				\begin{table}
					\small
					\begin{center}
						\caption{Comparison of 5G and 6G wireless systems}
						\label{tab:comparison2}
						\begin{tabular}{|p{2.8cm}| p{1.5cm} |p{2.1cm} |} 
							\hline
							& 5G & 6G \\ [0.5ex] 
							\hline\hline
							Peak data rates& 20 Gbps & 1 Tbps  \\ 
							\hline
							Experienced data rates& 0.1 Gbps & 1 Gbps  \\ 
							\hline
							End-to-end latency& 1 ms & 0.1 ms\\
							\hline
							Reliability  & $10^{-5}$&  ${10^{-9}}$ \\
							\hline
							Frequency bands & sub-6GHz, mmWave & sub-6GHz, mmWave, THz\\
							\hline
						\end{tabular}\\
						\vspace{4pt}
					\end{center}
				\end{table}

				%\begin{table}[t]
				%	\small
				%	\begin{center}
					%		\caption{Key technologies for metaverse wireless systems.}
					%		\label{tab:relatedwork}
					%		\begin{tabular}{|p{2cm}|p{5cm} |} 
						%			\hline
						%			Technologies & References \\ [0.5ex] 
						%			\hline
						%			Communication& Terahertz Communications \cite{akyildiz2022terahertz}; Intelligent Reconfigurable Surfaces \cite{liaskos2022software,liaskos2018new}; Massive Ultra-Reliable Low-Latency Communications \cite{saad2019vision}.\\
						%			\hline
						%			Networking & Space-air-ground-sea integrated networks\cite{tang2022roadmap}; New IP \cite{li2020new}; tactile Internet \cite{minopoulos2022opportunities};Edge intelligence \cite{lim2022realizing}; Over-the-top (OTT) multimedia streaming \cite{liu2021point}. \\
						%			\hline
						%			Caching & Edge caching \cite{van2022edge}; Joint design of caching, communication, and computing \cite{cai2022joint}. \\
						%			\hline
						%			Computation & Edge computing/distributed cloud \cite{tang2022roadmap,cai2022joint};  \\
						%			\hline
						%		\end{tabular}\\
					%		\vspace{4pt}
					%	\end{center}
				%\end{table}
				\subsection{Wireless communication}
				\label{sec:wc}
				To facilitate mobility, wireless communication is a critical technology to connect mulsemedia sensors and actuators. In this subsection, we discuss the physical layer challenges and potential solutions. To achieve low latency and high data rates, terahertz communication using frequencies from 0.1 to 10 THz can provide data rates of several terabits per second (Tbps), which is an important advancement of 6G wireless systems \cite{akyildiz2022terahertz}. Such a solution can be used to establish high-speed networks for mulsemedia sensors and displays. The range of communication can span several meters to tens of meters, depending on the location of mulsemedia devices. Accurate terahertz band channel models are necessary to enable adaptive wireless resource allocation. Additionally, the physical layer and data link layer communication protocols need to be reinvestigated and jointly designed by considering the specific mulsemedia applications. To prolong the operation time of wireless mulsemedia devices and avoid frequent recharging, new signal waveforms and media access technologies have to be developed in the terahertz band to improve energy efficiency. 
				
				Ubiquitous connectivity is required since users need to get access to mulsemedia content from any location. The space-air-ground-sea integrated networking is an essential technology to achieve this goal \cite{tang2022roadmap}. Space satellites and Unmanned Aerial Vehicles (UAVs) enable larger coverage that can provide wireless connectivity also in isolated areas without terrestrial wireless infrastructures.    
				
				For mulsemedia users with XR and HTC devices who have strong interactions with the virtual world, their activities can incur various body motions. Since terahertz wireless signals are highly directional, the alignment between terahertz transceivers is challenging due to the required precision. Also, the motion of the human body can affect the propagation of wireless signals. To this end, the Reconfigurable Intelligent Surfaces (RIS) can be leveraged to form a smart wireless environment to adaptively reflect and reroute wireless signals to ensure that the wireless link can be smoothly maintained \cite{liaskos2022software}\cite{liu2021learning}.   
				
				Semantic communication will play an important role in mulsemedia communication. {{Instead of transmitting information bits, semantic communication uses machine learning/deep learning tools to extract underlying meaning in text, images, audio, and video \cite{xie2021deep,weng2021semantic,xie2021task}.} Existing research has shown that semantic communication can be more reliable in the presence of noises and efficient in terms of delivering information \cite{xie2021deep,weng2021semantic}. In learning-based semantic communications, typically different neural network modules related to source encoding, channel encoding, modulation, channel decoding and source decoding are jointly optimized according to a loss function tailored to the considered media format. In \cite{bo2022learning}, a semantic communication approach is presented in which a predefined Binary Phase-Shift Keying (BPSK) constellation is employed for modulation. It should be noted that in most other semantic communication schemes proposed in the literature, impractical analog constellation symbols are generated and transmitted in order to avoid non-differentiable functions which cannot be trained with existing algorithms. In \cite{bo2022learning}, this problem is addressed by proposing a neural encoding network generating likelihoods of constellation points which is combined with a random coding-based modulator.  The corresponding results demonstrate that existing digital modulation methods in semantic communications and neural network-based analog modulation, respectively, can be outperformed by the proposed scheme.  More research in this direction is required in order to design efficient and practically relevant joint modulation and coding schemes for mulsemedia semantic communications with arbitrary discrete symbol constellations.

					Although mulsemedia communication systems leverage multiple human senses, the generated multisensory media have strong semantic correlations because they represent the same perception of the physical world. Multimodal machine learning \cite{baltruvsaitis2018multimodal} is an extensively studied machine learning area that is strongly related to mulsemedia semantic communication. Existing work mainly studies the semantic correlations among text, audio, image, and video. By using deep generative models, we can generate an unobserved media format with one or several available media formats, e.g., from text and image to videos \cite{wang2020imaginator}. This approach can be extended to mulsemedia communication and generate multisensory media that cannot be easily captured by sensors.

					\subsection{Integrated communication and sensing}
					Wireless sensing can reduce the weight and power consumption of wearable mulsemedia devices, such as XR HMDs, by using external access points, base stations, or ambient wireless signals. Wireless sensing can enhance the recording and display quality of mulsemedia content. Motion sensing is essential for mulsemedia communication systems, as it enables adaptive rendering of mulsemedia content. For example, in XR systems the communication data rate and virtual content can be optimally selected based on the FoV \cite{he2018joint}, which is enabled by FoV sensing. Also, prediction of a user's motion based on previous sensing knowledge can reduce the end-to-end latency. Machine learning can be used to predict the FoV based on the user's motion and preselect the optimal communication parameters. 
					
					6G wireless systems can leverage terahertz signals for motion sensing, as they have a high sensing resolution due to their short wavelength \cite{akyildiz2022terahertz}. Terahertz beamforming using massive MIMO can track users and sense their motion with steerable beams. Terahertz sensing can also use active radar systems with dedicated transmitters and receivers or ambient systems that analyze and detect changes in the propagation and distribution of terahertz signals of the background environment. While terahertz radar has been used to classify gestures \cite{wang2020negative}, its accuracy can still be improved. 
					
					It should be noted that sensing should be codesigned with communication in 6G wireless systems. Joint Communication and Sensing (JCAS) is an integral part of 6G networks \cite{fang2022joint}. The massive MIMO systems, RIS, mmWave signals, terahertz signals, and many other wireless technologies provide high resolution of sensing. The coexistence of sensing and communication poses new research problems. New algorithms and protocols are desirable for mulsemedia applications. 
					
					\subsection{Mulsemedia streaming}
					Mulsemedia communication requires unprecedented high throughput (several Tbps) to meet the demands of multiple users. Such high throughput necessitates either a new network architecture design or improvements to the existing one. The New IP protocol introduced in \cite{li2020new} expands existing Internet protocols to add more services and mechanisms, such as adaptive packet dropping and guaranteed latency. {{Various network support technologies are essential for mulsemedia streaming, such as in-network caching, bandwidth saving, and adaptive streaming \cite{garcia2018network}.} To achieve high data rates and low end-to-end latency, simply increasing wireless communication capacity is not effective. Existing studies of 5G wireless systems have shown that the wireline network is the bottleneck for reducing end-to-end latency \cite{xu2020understanding}. The edge and core networks also need updating to provide new services, such as precision networks that can deliver data packets at predefined times with guaranteed variances. Additionally, edge caching can efficiently reduce network traffic. For example, caching frequently requested information such as the floorplan, store information, and product information in local edge servers for AR and MR users in a shopping mall can optimize storage and communication latency. 
						
						Besides general network upgrade, mulsemedia streaming can be specifically developed by adding additional sensing media to existing multimedia streaming protocols. The hidden visual channel communication or hidden acoustic channel communication can be used to deliver mulsemedia \cite{xu2022lsync}. In \cite{xu2022lsync}, the metadata is modulated using Chirp Spread Spectrum (CSS) in the format of acoustic signals. The Advanced Audio Coding (AAC) encodes the metadata together with audio signals. The destination decoder is carefully designed to extract the metadata without affecting audio signal quality. Similar techniques can be used to insert the metadata into visual channels for video encoding. The metadata can include the mulsemedia information such as the SEM in MPEG-V. Although this approach is low-cost and low-complexity, the amount of mulsemedia data that can be transmitted is limited. The metadata can be transmitted with 156.25 bps in \cite{xu2022lsync}. It remains a challenge to support interactive mulsemedia applications.

						\subsection{Mulsemedia synchronization}
						
						In Fig.~\ref{fig:tsn}, we show the required data rates, latency, and packet loss rate for five basic sensory media. The high-quality haptic media can be used for teleoperations and the fair experience requirements are simple haptic signals that can be used for gaming with haptic controllers. The olfactory and gustatory requirements are suggested values based on other sensory media since the products are not widely available. As we can see, the bottleneck of mulsemedia data rates is video and all other media have much smaller data rates. However, it is not trivial to include more senses due to the complex synchronization problem \cite{yuan2015perceived,ademoye2009synchronization}.   
						
						Mulsemedia communication requires strict synchronization of multiple sensory data flows, which includes (1) intra-media synchronization and (2) inter-media synchronization. The intra-media synchronization reorganizes received data packets of single sensory media, while the inter-media synchronization coordinates the display of multiple sensory media, as shown in Fig.~\ref{fig:sync}. Similar problems have been already investigated in multimedia (audio and video) communication in the 1990s \cite{akyildiz1996multimedia,rangan1993communication,steinmetz1990synchronization,little1991multimedia}. In particular, the lip-synch problem (jitter problem) was investigated by many researchers \cite{akyildiz1996multimedia,ishibashi1997group}. However, in mulsemedia communication the number of sensory media to be synchronized are significantly increased and, accordingly, there are many more additional research challenges. 
						
						\begin{figure}[t]
							\centering
							\includegraphics[width=0.49\textwidth]{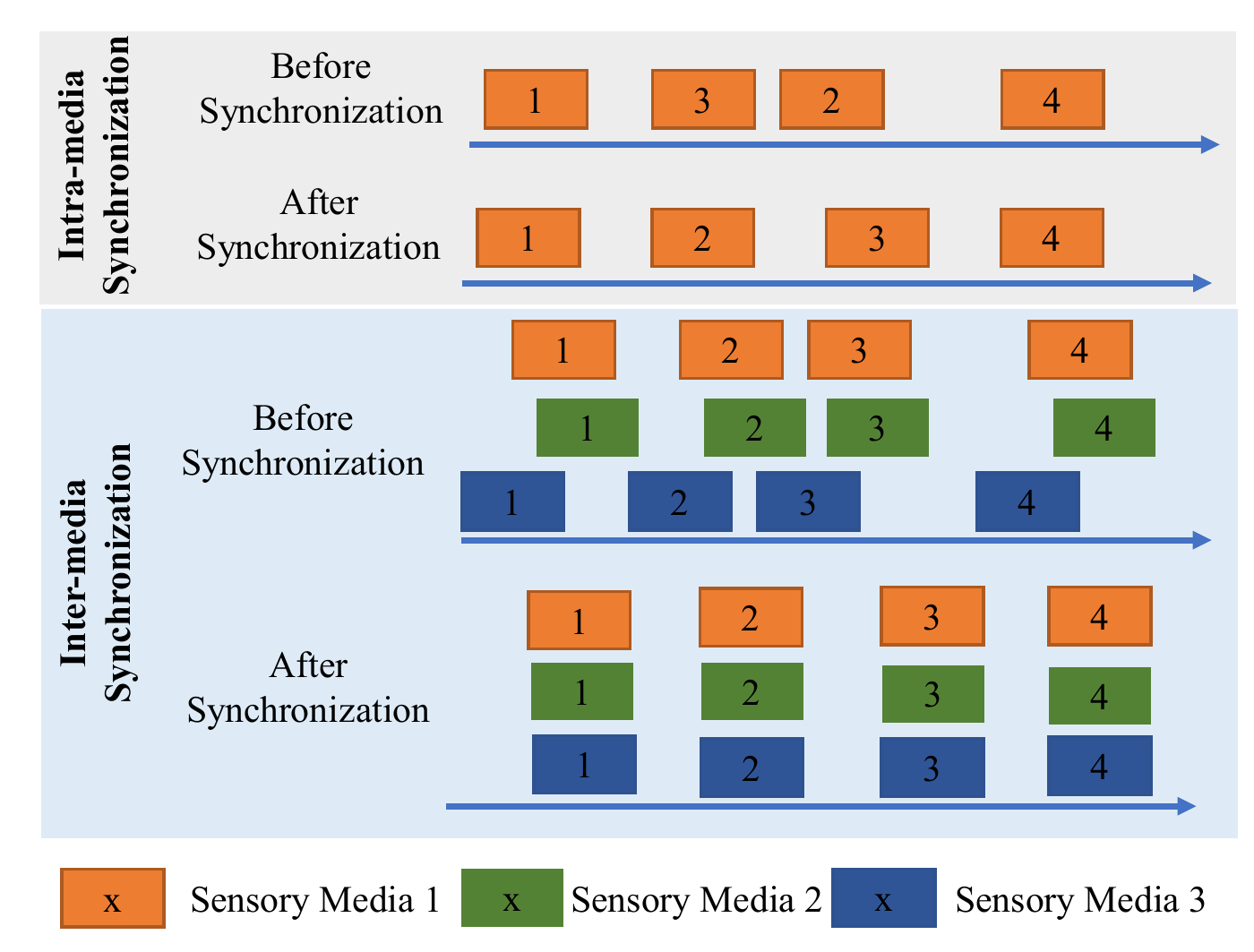}
							\vspace{-5pt}
							\caption{Illustration of intra-media and inter-media synchronization. }
							\label{fig:sync}
							\vspace{-0pt}
						\end{figure}
						
						Note that if the buffering solution is used then the end-to-end latency can be affected which is a strict QoE in mulsemedia communication. {{First, the application can reorder packets after receiving and buffering them. However, this requires a sufficient buffer size and the application needs to be delay-tolerant.}} Another possible solution could be using traffic shaping solutions \cite{fu2022survey} such as leaky bucket and token bucket algorithms from existing literature. Despite improvements in communication data rates and network latency reduction in previous generations of wireless networks, to the best of our knowledge there is currently no wireless technology capable of delivering data packets with bounded end-to-end latency and synchronized, coordinated data flows. This differs from the requirement of low latency, which would only guarantee that a large percentage of data packets meet the deadline with a certain number of outliers. As a result, new technologies are required for 6G wireless systems to meet the demands of mulsemedia communication.

						%Time-Sensitive Networking (TSN) is a collection of protocols designed by IEEE 802.1 working group for deterministic Ethernet networks. It enables users to define the end-to-end latency, and the network scheduler optimizes the allocation of network resources to configure the network and meet these requirements. This results in a nearly constant end-to-end latency in TSN. TSN switches utilize queues and gating control policies to ensure that network traffic is delivered with a consistent end-to-end latency. TSN, along with underlying technologies like traffic shaping, frame preemption, frame replication, and cut-through switching, can be leveraged to provide better synchronization and end-to-end latency control for mulsemedia communication. TSN has recently been extended to wireless networks \cite{cavalcanti2019extending}. Designing wireless TSN remains an open problem due to the unreliable nature of the wireless channel.
						
						\subsection{Artificial intelligence}
						Besides semantic communication discussed in Section~\ref{sec:wc}, artificial intelligence, especially basic machine learning and deep learning models, can be used in many other areas of mulsemedia communication systems to improve system performance. To enable widespread adoption of mulsemedia, it is essential to explore immersive video and hologram generation using more accessible devices such as smartphones. However, current technologies for generating and rendering immersive videos and holograms require numerous fixed camera sensors and sophisticated processing tools that are not widely available. Moreover, the generation, encoding/decoding, and rendering of 3D visual information involves several processing steps that can cause significant delays. Therefore, it is expected that mobile edge computing will play a crucial role in providing additional computation capability. Edge intelligence techniques \cite{xu2021edge}, such as intelligent data compression and intelligent offloading of high-complexity tasks to the cloud, can help accomplish computational tasks on time. To ensure that end-to-end latency requirements of high-interactive applications are met, lightweight computation algorithms should be designed. It is also essential to jointly design computation and communication components, considering multiple sources of latency.
						
						The recently developed Large Language Model (LLM) can also facilitate the design and commercialization of mulsemedia communication systems. It has been shown that LLM can be used to design chips \cite{blocklove2023chip} and this idea can be extended to communication systems design \cite{bariah2023large}. Multimodal large models can be trained on data collected from existing data communication systems which can be used to simplify the design and management of existing wireless and wired communication networks. Mulsemedia communication system designers can interact with data communication LLM and obtain suggestions and even solutions. Besides interactive data communication LLM, a dedicated Large Data Communication Model (LDCM) can be trained for autonomous driving networking, which can self-heal, self-diagonose, self-monitor, self-configure, and self-report. Mulsemedia communication users can send a request given required sensory media, bandwidth, end-to-end latency, etc. The LDCM will autonomously schedule resources and connect the source user with the destination. The research in zero touch networks, autonomous driving networks, and self driving networks has the potential to realize this vision \cite{rossi2022landing,benzaid2020ai}.

						\subsection{QoE modeling, management, and optimization}
						QoE measures the degree of delight or annoyance of users. It can be obtained by using subjective measures, where users rate their experience after using the service or application, and objective measures, where physiological signals such as accelerometers, video recordings, and biosignals are used to evaluate users' biofeedback and responses \cite{yuan2014user,da2022ongoing}. A model of QoE is essential to design optimal mulsemedia communication systems. The QoE of mulsemedia can be modeled as a linear or nonlinear combination of each individual sense's QoE. The coefficients for this combination can be determined by machine learning models trained on data collected from system configurations and user ratings. Also, it is interesting to investigate the impact of physical layer parameters such as transmit power and power of interference from other users, and physical parameters influencing the behavior of the THz channel like the water vapor percentage, on the QoE of users. Such study has been conducted in \cite{chaccour2022can} for VR users, indicating that using frequencies with low molecular absorption and the presence of a line-of-sight path are crucial. As mentioned in \cite{chaccour2022can}, a separate analysis should be performed for holographic/mulsemedia communications which have different QoE requirements. For example, five senses have to arrive synchronized, i.e., the instantaneous delay stemming from each sense is identical and extremely low. Furthermore, it would be of great interest to extend the model in \cite{chaccour2022can} such that it reflects the dependency of QoE on the detailed waveform of transmission, and to employ such a tool for waveform optimization. With a QoE model, mulsemedia streaming can be optimally designed by optimizing communication systems, networking protocols, and available resources and adaptively streaming mulsemedia.

						\section{Conclusion} 
						Mulsemedia technology has the potential to provide users with highly immersive experiences that engage multiple senses beyond just sight and hearing. By transmitting multisensory content through both wired and wireless networks, mulsemedia communication represents an exciting and rapidly evolving field of study. In this paper, we provide an overview of the background, history, and essential encoding, decoding, and networking technologies that underpin this emerging field. We introduce the two major mulsemedia applications, namely, {{extended reality (XR) and Holographic-Type Communication (HTC)}}. Additionally, we highlight potential research directions for mulsemedia communication in the context of 6G wireless systems, which have the potential to further enhance the capabilities and performance of this exciting technology.

			\section*{Acknowledgement}
			\label{sec:ackn}
			The work of Ian F. Akyildiz was supported by the Alexander von Humboldt Prize for his research stay at the Friedrich-Alexander-Universität Erlangen-Nürnberg, Germany in 2023.

\bibliographystyle{IEEEtran}
\bibliography{ref2}

\end{document}